\documentclass[twocolumn,twoside,slac]{revtex4}
\usepackage{graphicx}
\usepackage{fancyhdr}
\usepackage{psfig,epsfig}
\newcommand{\Pom}{I\!P}
\newcommand{\bea}{\begin{eqnarray}}
\newcommand{\eea}{\end{eqnarray}}
\newcommand{\FLUKA}{{\sc FLUKA}}

\newcommand{\DPMJET}{{\sc DPMJET}}
\newcommand{\RQMD}{{\sc RQMD}}
\newcommand{\bino}[2]{\left ( \begin{array}{c} #1\\#2 \end{array} \right )}
\newcommand{\bes}{\begin{eqnarray*}}
\newcommand{\ees}{\end{eqnarray*}}
\newcommand{\der}{\mathrm{d}}

\begin{document}

\title{The physics models of FLUKA: status and recent developments}


\author{A.~Fass\`o\footnote{on leave from GSI, Darmstadt, Germany
(Present address: SLAC, P.O. Box 4349, MS 48, Stanford, CA 94309, USA)},
A.~Ferrari\footnote{on leave from INFN, Sezione di Milano,
via Celoria 16, I-20133, Milan, Italy}, S.~Roesler}
\affiliation{CERN, Geneva, CH-1211 Switzerland}
\author{P.R.~Sala$^\dagger$}
\affiliation{ETH Z\"urich, CH-8093 Z\"urich, Switzerland}
\author{F.~Ballarini, A.~Ottolenghi}
\affiliation{Universit\`a degli Studi di Pavia and INFN, via Bassi 6, I-27100,
Pavia, Italy}
\author{G.~Battistoni, F.~Cerutti, E.~Gadioli, M.V.~ Garzelli}
\affiliation{Universit\`a degli Studi di Milano and INFN, via Celoria 16,
I-20133, Milan, Italy}
\author{A.~Empl}
\affiliation{Houston University, Texas, USA}
\author{J.~Ranft}
\affiliation{Siegen University, Germany}

\begin{abstract}
A description of the intermediate and high energy hadronic interaction
models used in the \FLUKA{} code is given.
Benchmarking against experimental data
is also reported in order to validate the model performances.
Finally the most recent developments and perspectives for nucleus--nucleus
interactions are described together with some comparisons with
experimental data.
\end{abstract}

\maketitle

\thispagestyle{fancy}


\section{Generalities}
\FLUKA{}~\cite{FLUKA,FLUKA2,Trieste,Stresa} is a multipurpose transport
Monte Carlo code, able to treat hadron--hadron, hadron--nucleus, neutrino,
electromagnetic,
and $\mu$ interactions up to 10000~TeV. Charged particle transport (handled in
magnetic field too) includes all relevant processes \cite{FLUKA}.
About nucleus--nucleus
collisions, since ion--ion nuclear interactions were not yet treated in \FLUKA,
past results have been obtained in the superposition model approximation, where
primary nuclei (0--10000~TeV/A) were split into nucleons before interacting.
With the integration of ion interaction codes (see Section \ref{SCAAcoll})
and the cross section parameterization, this approximation is now obsolete.

\FLUKA{} is based, as far as possible, on original and well tested microscopic
models. Due to this ``microscopic'' approach to hadronic interaction modelling,
each step is self--consistent and has solid physical bases.
Performances are optimized comparing with particle production data at
single interaction level. No tuning whatsoever is performed on ``integral'' data,
such as calorimeter resolutions, thick target yields, etc.
Therefore, final predictions are obtained with a minimal set of free parameters,
fixed for all energies and target/projectile combinations.
Results in complex cases as well as scaling laws and properties come forth
naturally from the underlying physical models and the basic conservation
laws are fulfilled {\it a priori}.

\section{Hadron--nucleon interaction models \label{SSChn}}
A comprehensive understanding of hadron--nucleon (h--N) interactions over a
wide energy range is of course a basic ingredient for a sound description of
hadron--nucleus ones.
Elastic, charge exchange and strangeness exchange reactions
are described as far as possible by phase--shift analysis and/or fits
of experimental differential data. Standard eikonal
approximations are often used at high energies.

At the low energy end (below 100~MeV) the p--p and p--n cross sections
are rapidly increasing with decreasing energy. The n--p and the p--p cross 
sections differ by about a factor three at the lowest energies, as expected 
on the basis of symmetry and isospin considerations, while at high energies 
they tend to be equal.

The total cross section for the two isospin components present in the
nucleon--nucleon amplitude is given by:
\bea
     \sigma_1&=&\sigma_{pp} \nonumber \\
     \sigma_0&=&2\sigma_{np}-\sigma_{pp}\, . \nonumber
\eea

The same decomposition can be shown to apply for the elastic and the
reaction cross sections too.

The cross section for pion--nucleon scattering is dominated by the
existence of several direct resonances, the most prominent one being the
$\Delta(1232)$.
Given the pion isotopic spin ($T=1$), the three $\pi$ charge states
correspond to the three values of $T_z$. Thus, in the pion-nucleon system two
values of $T$ are allowed : $T= \frac{1}{2}$ and $T= \frac{3}{2}$, and two
independent scattering amplitudes, $A_\frac{1}{2}$ and $A_\frac{3}{2}$,
enter in the cross sections.
Using Clebsch-Gordan coefficients all differential cross sections
can be derived from
the three measured ones: $ \sigma  \left(\pi^+ p \rightarrow \pi^+
p\right)$,  $ \sigma  \left(\pi^- p \rightarrow \pi^- p\right)$,
and $\sigma  \left(\pi^- p \rightarrow \pi^0 n\right)$.

As soon as particle production is concerned (inelastic hadron-nucleon interactions),
the description becomes immediately more complex.
Two families  of models are adopted, depending on the projectile energy:
those based on individual resonance production and decays, which cover
the energy range up to 3--5~GeV, and
those based on quark/parton string models, which can provide
reliable results up to several tens of TeV.
%
\subsection{h-N interactions at intermediate energies \label{SSSChninter}}
The inelastic channel with the lowest threshold, single pion production, opens
already around 290~MeV in nucleon-nucleon interactions,
and becomes important above 700~MeV. In pion-nucleon interactions
the production threshold is as low as 170~MeV.
Both reactions
are normally described
in the framework of the isobar model: all reactions
proceed through an intermediate state containing at least one
resonance. There are two main classes of reactions, those which form a
resonant intermediate state (possible in $\pi$-nucleon reactions) and those
which contain two particles in the intermediate state. The former exhibit
bumps in the cross sections corresponding to the energy of the formed
resonance. Examples are reported below:
\bea
N_1 + N _2 & \rightarrow & N_1^\prime + \Delta(1232)\ \rightarrow
N_1^\prime + N_2^\prime +\pi \nonumber \\
\pi + N & \rightarrow &\ \Delta (1600)\ \rightarrow \pi^\prime +
\Delta(1232) \rightarrow \nonumber\\
& \rightarrow & \pi^\prime + \pi^{\prime\prime} +N ^\prime \nonumber\\
N_1 + N _2 & \rightarrow & \Delta_1(1232) + \Delta_2(1232) \rightarrow
\nonumber\\
& \rightarrow & N_1^\prime + \pi_1 + N_2^\prime +\pi_2 \nonumber
\eea

Partial cross sections can be obtained from one--boson exchange theories and/or
folding of Breit--Wigner with matrix elements fixed by N--N
scattering or experimental data.
Resonance energies, widths, cross sections, and branching
ratios are extracted from data and conservation laws, whenever possible,
making explicit use of spin and isospin relations.
They can be also inferred from inclusive cross sections when needed.

For a discussion of resonance production, see for
example~\cite{RES1,RES2,RES3}.
\subsection{Inelastic h-N interactions at high energies
\label{SSSChn high}}
\begin{figure}[htb]
\begin{center}
\epsfig{file=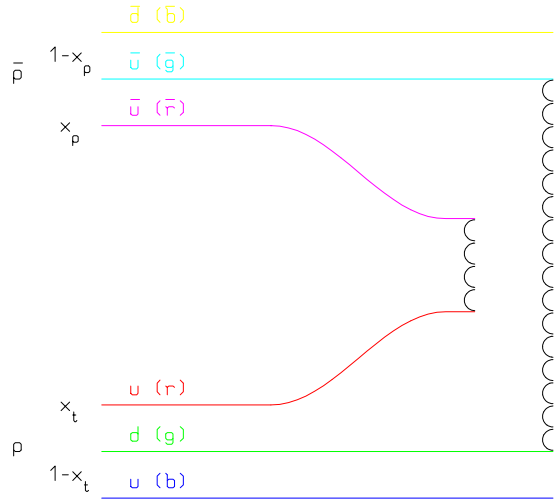,width=0.45\textwidth,bbllx=25pt,%
bblly=50pt,bburx=560pt,bbury=560pt}
\caption{Leading two-chain diagram in DPM for $\bar p-p$ scattering.
The colour (r$\rightarrow$red, b$\rightarrow$blue, g$\rightarrow$green, $\bar
\mathrm{r}\rightarrow$antired, $\bar \mathrm{b}\rightarrow$antiblue and $\bar
\mathrm{g}\rightarrow$antigreen) and quark combination shown in the figure
is just one of the allowed possibilities.
\label{fig:pbarpchain}
}
\end{center}
\end{figure}
\begin{figure}[htb]
\begin{center}
\epsfig{file=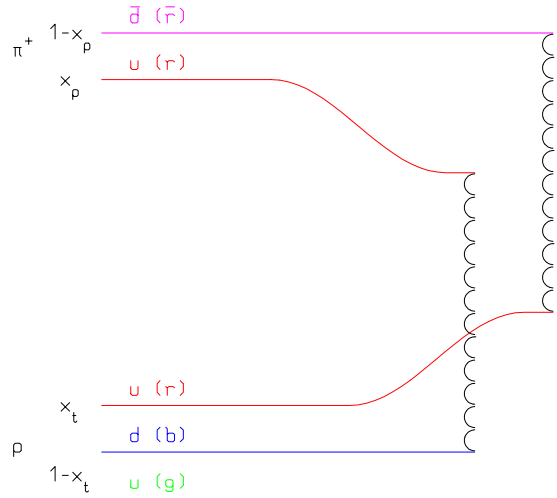,width=0.45\textwidth,bbllx=25pt,%
bblly=50pt,bburx=560pt,bbury=560pt}
\caption{One of leading two-chain diagrams in DPM for $\pi^+-p$
scattering. The colour and quark combination
shown in the figure is just one of the allowed ones.
 \label{fig:pippchain}
}
\end{center}
\end{figure}
\begin{figure}[bth]
\begin{center}
\epsfig{file=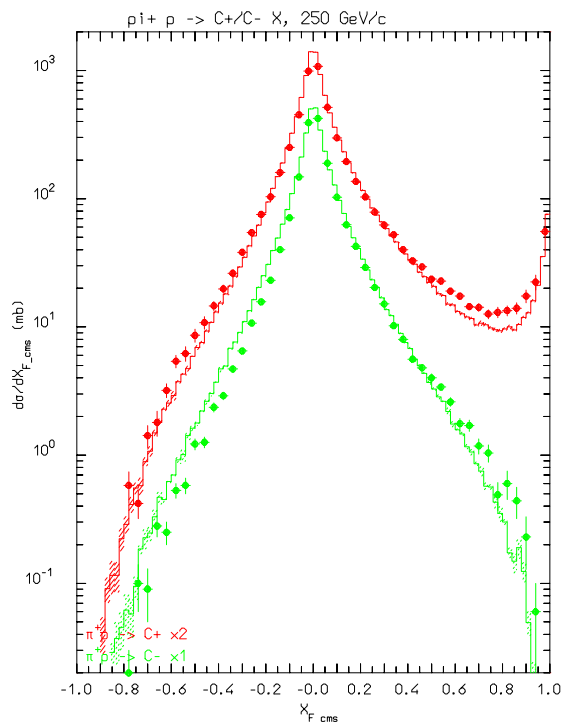,width=0.45\textwidth%
,bbllx=60pt,bblly=100pt,bburx=590pt,bbury=740pt}
\caption{
Feynman $x^*_F$ spectra of positive and negative
particles from ($\pi^+$,~p) at 250~GeV/c~.
Exp.~data (symbols) from~\protect\cite{NA22}.
 \label{fig:hh250x}
}
\end{center}
\end{figure}
\begin{figure}[bth]
\begin{center}
\epsfig{file=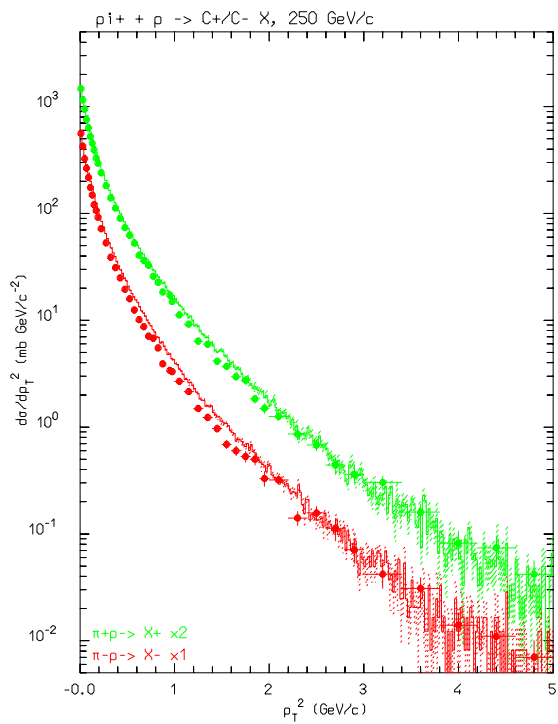,width=0.45\textwidth%
,bbllx=60pt,bblly=100pt,bburx=590pt,bbury=740pt}
\caption{
Transverse momentum ($p_t$) spectra of positive and negative
particles from ($\pi^+$,~p) at 250~GeV/c~.
Exp.~data (symbols) from~\protect\cite{NA22}.
 \label{fig:hh250pt}
}
\end{center}
\end{figure}
As soon as the projectile energy exceeds a few GeV, the description in
terms of resonance prodution and decay becomes more and more
difficult. The number of resonances which should be considered grows
exponentially and their properties are often poorly known. Furthermore, the
assumption of one or two resonance creation is unable to reproduce the
experimental finding that most of the particle production at high energies
occurs neither in the projectile nor in the target fragmentation region, but rather
in the central region, for small values of Feynman~$x$ variable. Different
models, based directly on quark degrees of freedom, must be introduced.

The features of ``soft''
interactions (low-$p_T$ interactions) cannot be derived from the QCD
Lagrangian, because the large value taken by the
running coupling constant prevents the use of perturbation theory.
Models
based on interacting strings have emerged as a powerful tool in understanding
QCD at the soft hadronic scale, that is in the non-perturbative regime.
An interacting string theory naturally leads to a topological expansion.
The Dual Parton Model (DPM)~\cite{DPM} is one of these models
and it
is built introducing partonic ideas into a topological expansion which
explicitly incorporates the
constraints of duality and unitarity, typical of Regge's theory.
In DPM hadrons are considered as open strings with quarks,
antiquarks or diquarks sitting at the ends;
mesons (colourless combination of a quark and an antiquark $q\bar q$)
are strings with their valence quark and antiquark at the
ends.
At sufficiently high energies the leading term in the interactions
corresponds to a Pomeron ($\Pom$) exchange (a closed string exchange),
which has a cylinder topology.
When an unitarity cut is applied to the cylindrical Pomeron, two hadronic
chains are left as the sources of particle production.
As a consequence of colour
exchange in the interaction, each colliding hadron splits into two
coloured system, one carrying colour charge $c$ and the other $\bar c$.
The
system with colour charge $c$ ($\bar c$) of one hadron combines with the
system of complementary colour of the other hadron, to
form two colour neutral chains. These chains appear as two back-to-back
jets in their own centre-of-mass systems.

The exact way of building up these chains depends on the nature of the
projectile--target combination (baryon--baryon, meson--baryon,
antibaryon--baryon, meson--meson): examples are shown in
figs.~\ref{fig:pbarpchain} and~\ref{fig:pippchain}.
Further details can be found in the original DPM references~\cite{DPM} or
in \cite{Trieste}.

The chains produced in an interaction are then hadronized. DPM gives no
prescriptions on this stage of the reaction. All the available chain
hadronization models, however, rely on the same basic assumptions, the most
important one being {\it chain universality}, that is chain hadronization
does not depend on the particular process which originated the chain, and
until the chain energy is much larger than the mass of the hadrons to be
produced, the fragmentation functions (which describe the momentum fraction
carried by each hadron) are the same.
As a consequence, fragmentation functions can in principle be derived
from hard processes and $e^+e^-$ data and the same functions and (few)
parameters should be valid for all reactions and energies;
actually mass and threshold effects
are non-negligible at the typical chain energies involved in hadron-nucleus
reactions. Transverse momentum is usually added according to uncertainty
considerations.
The examples in figs.~\ref{fig:hh250x} and~\ref{fig:hh250pt} show the ability
of the \FLUKA{} model, based on DPM,
to reproduce the features of particle production; further examples can be
found in~\cite{FLUKA,Trieste}.

\section{Main steps of hadron--nucleus interactions \label{SChaintro}}
High energy hadron--nucleus (h--A) interactions can be schematically
described as a sequence of the following steps:
\begin{itemize}
\item Glauber--Gribov cascade
\item (Generalized) IntraNuclear cascade ((G)INC)
\item Preequilibrium emission
\item Evaporation/Fragmentation/Fission and final deexcitation
\end{itemize}

\subsection{The Glauber-Gribov cascade and the formation zone
\label{SSCglauber}}
\begin{figure}[tbh]
\begin{center}
\epsfig{file=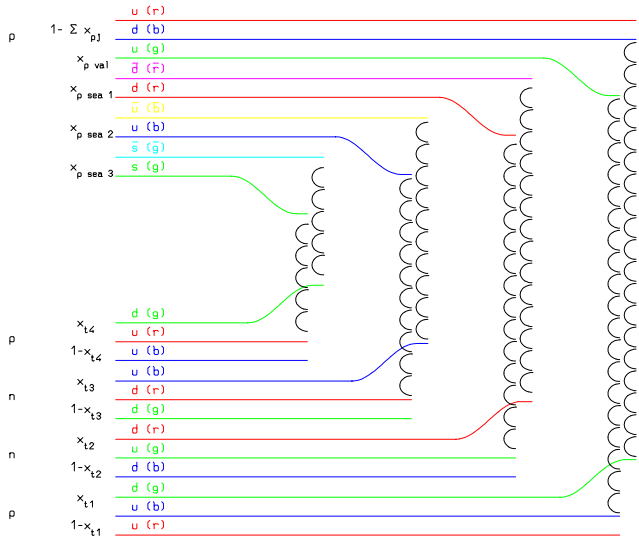,width=0.45\textwidth%
,bbllx=60pt,bblly=20pt,bburx=550pt,bbury=520pt}
\caption{
Leading two-chain diagrams in DPM for $p-A$ Glauber scattering with 4
collisions. The colour and quark combinations shown
in the figure are just one of the allowed possibilities.
\label{fig:pglauchain}
}
\end{center}
\end{figure}
\begin{figure}[tbh]
\begin{center}
\epsfig{file=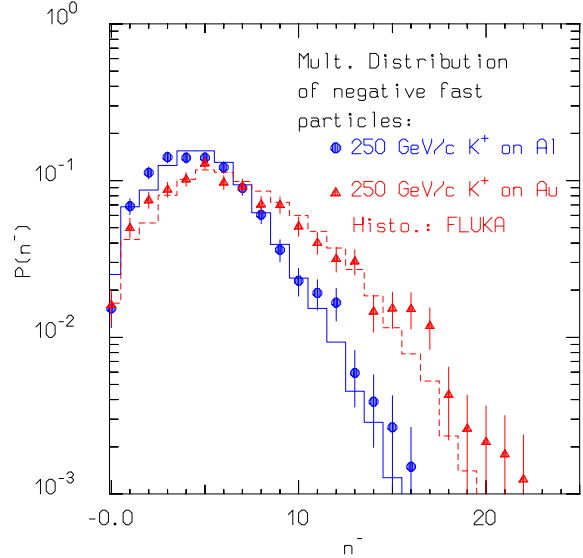,width=0.45\textwidth%
,bbllx=60pt,bblly=30pt,bburx=590pt,bbury=640pt}
\caption{
Multiplicity distribution of negative shower particles for
250~GeV/c K$^+$ on aluminium and gold targets.
Data from~\protect\cite{NA22ALAUM}.
 \label{fig:kno250}
}
\end{center}
\end{figure}
\begin{figure}[htb]
\begin{center}
\epsfig{file=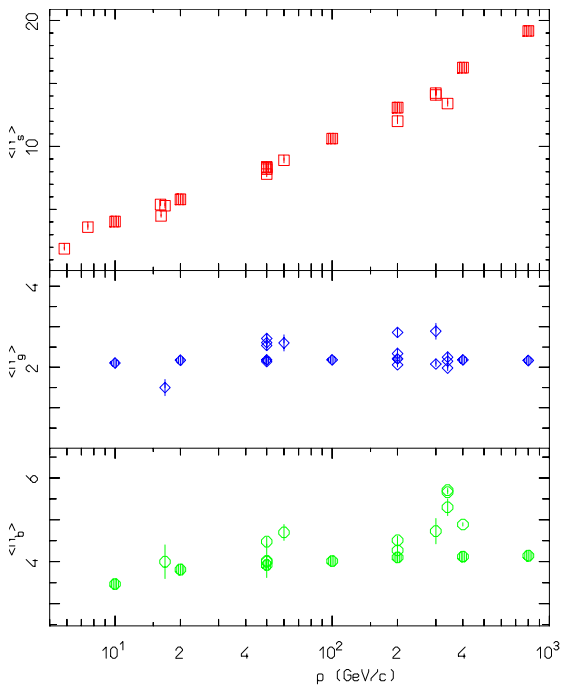,width=0.45\textwidth%
,bbllx=60pt,bblly=50pt,bburx=590pt,bbury=680pt,clip=}
\vfill
\caption{
Shower, grey, and black tracks multiplicities for $\pi^-$
on emulsion, as a function of the projectile
momentum. Open symbols are experimental data from various sources,
full symbols are \FLUKA{} results.
 \label{fig:ppiem1}
}
\end{center}
\end{figure}
The Glauber formalism~\cite{GLAUB,GLAUB2} provides a powerful and elegant
method to derive elastic, quasi-elastic and absorption h--A cross sections from
the free h--N cross section and the nuclear ground state
{\it only}.
Inelastic interactions are equivalent to  multiple interactions of the
projectile with $\nu$
target nucleons. The number of such ``primary''
interactions follows a binomial distribution
 (at a given impact parameter, $b$):
\bea
P_{r\nu\,} (b) & \equiv & \bino{A}{\nu} P_r^{\,\nu}(b)\,\left[1-P_r(b)
\right]^{A-\nu}\nonumber
\eea
where $P_{r} (b) \equiv \sigma_{hNr\,} T_{r}(b) $, and
$T_{r}(b)$ is the  profile function (folding
of nuclear density and scattering profiles along the path).
On average:
\bes
<\nu> &=& \frac{Z\sigma_{hp\,r} + N\sigma_{hnr}}{\sigma_{hA\,abs}}
\ees
\bes
\sigma_{hA\,abs}(s)& = &
\int{\der ^2\vec b\left [ 1 - \left(1 -
\sigma_{hNr}(s)\, T_{r}(b) \right)^A \right] }\, .
\ees

The Glauber-Gribov model~\cite{GRIBOV,GRIBOV2,GRIBOV3} represents the
diagram interpretation of the Glauber cascade.
The $\nu$ interactions of the projectile originate
2$\nu$ chains, out of which
2 chains (valence-valence chains) struck between the projectile and
target valence (di)quarks,
$2(\nu-1)$ chains (sea-valence chains) between projectile sea $q-\bar q$
and target valence (di)quarks.

A pictorial example of the chain building process is depicted in
fig.~\ref{fig:pglauchain} for p--A: similar diagrams apply to $\pi$--A and
$\bar p$ --A respectively.

The distribution of the projectile energy among many chains naturally
softens the energy distributions of reaction products
and boosts the multiplicity with respect to hadron-hadron interactions.
The building up of the multiplicity distribution
from the multiple collisions can be appreciated from fig.~\ref{fig:kno250},
where the multiplicity distribution for Al and Au targets at 250~GeV/c are
presented together.
In this way, the model accounts for the major A-dependent features
without any degree of freedom, except in the treatment of mass effects at
low energies.

The Fermi motion of the target nucleons must be included to obtain the
correct kinematics, in particular the smearing of p$_T$ distributions.
All nuclear effects on the secondaries
are accounted for by the subsequent (G)INC.

The {\it formation zone} concept is essential to understand the observed
reduction of the re-interaction probability with respect of the naive free
cross section assumption.
It can be understood as a ``materialization" time.
At high energies, the ``fast'' (from the emulsion language) particles
produced in the Glauber cascade have a high probability to materialize
already outside the
nucleus without triggering a secondary cascade. Only a small fraction of
the projectile energy is thus left available for the INC and the
evaporation.

The Glauber cascade and the formation zone act together in reaching a
regime where the ``slow'' part of the interaction
is almost independent of the particle energy.
This can be easily verified looking at charged particle average
multiplicities and multiplicity distributions as a function of energy
(fig.~\ref{fig:ppiem1}).
``Fast'' (or ``shower'') tracks (charged particles
with $\beta=\frac{v}{c}>0.7$), coming from the projectile primary interactions,
show the typical $\approx$~logarithmic increase observed for h--N
interactions. As shown already in
fig.~\ref{fig:kno250},
the average multiplicity and its variance are
directly related to the distribution of primary collisions as predicted by
the Glauber approach. Due to the very slow variation of h--N
cross section from a few GeV up to a few TeV, the Glauber cascade is
almost energy independent and the rise in the multiplicity of ``fast''
particles is related only to the increased multiplicity of the elementary
h--N interactions.

Due to the onset of formation zone effects, most of the hadrons produced
in the primary collisions escape from the nucleus without further
reinteractions. Further cascading only involves the slow fragments
produced in the target fragmentation region of each primary interaction, and
therefore it tends quickly to saturate with energy as the Glauber cascade
reaches its asymptotic regime. This trend is well reflected in the average
multiplicity (and multiplicity distribution) of ``gray'' tracks (charged
particles with $0.3<\beta<0.7$), which are mostly protons produced
in secondary collisions during the INC and preequilibrium
phases.

At the end of the cascading process, the residual excitation energy
is directly related to the number of primary and secondary collisions which
have taken place. Each collision is indeed leaving a ``hole'' in the Fermi sea
which carries an excitation energy related to its depth in the Fermi
sea.
Evaporation products, as well as residual excitation functions, should reach
an almost constant condition as soon as the Glauber mechanism and the
formation zone are fully developed. This can actually be verified by looking at
the production of ``black'' tracks (charged particles with $\beta<0.3$),
which are mostly evaporation products. The data reported in
fig.~\ref{fig:ppiem1}
do indeed demonstrate how
they saturate as well, and how this property is well reproduced on the
basis of the outlined assumptions.

\subsection{(Generalized) IntraNuclear cascade \label{SSCginc}}
At energies
high enough to consider coherent effects as corrections, a h--A
reaction can be described as a cascade of two-body interactions,
concerning the projectile and the reaction products. This is the mechanism
called IntraNuclear Cascade (INC). INC models were developed
already at the infancy of the computer era with great success in describing
the basic features of nuclear interactions in the 0.2-2~GeV range. Modern
INC models had to incorporate many more ideas and effects in order to
describe in a reasonable way reactions at higher and lower energies.
Despite particle trajectories are described classically, many quantistic
effects have to be incorporated in these (G)INC models, like
Pauli blocking, formation time, coherence length, nucleon antisymmetrization,
hard core nucleon correlations. A thorough description of the (G)INC model
used in \FLUKA{} can be found in~\cite{FLUKA,Trieste}.
\subsection{Preequilibrium \label{SSCpreeq}}
At energies lower than the $\pi$ production threshold
a variety of preequilibrium models have been developed~\cite{Gad92} following
two leading approaches:
the quantum-mechanical multistep model
and the exciton model.
The former has a very good theoretical background but is quite complex,
while the latter relies on statistical assumptions,
and it is simple and fast. Exciton-based models are often used in
Monte Carlo codes to join the INC stage of the reaction to the equilibrium one.

In the \FLUKA{} implementation
the INC step goes on until all nucleons are below a smooth threshold
around 50~MeV, {\it and} all particles but nucleons
(typically pions) have been emitted or absorbed. At the end of the INC
stage a few particles may have been emitted and the input configuration
for the preequilibrium stage
is characterized by the total number of protons and
neutrons,  by the number of particle-like excitons (nucleons
excited above the Fermi level), and of
hole-like excitons (holes created in the Fermi sea by the INC
interactions), and by the nuclear excitation energy and
momentum. All the above quantities can be derived by properly recording
what occurred during the INC stage.
The exciton formalism of \FLUKA{} follows that of M.~Blann
and coworkers~\cite{Blann71,Blann72,Blann83a,Blann83b}, with some
modifications detailed in~\cite{Trieste}.
\subsection{Evaporation, fission and nuclear break-up \label{SSCevap}}
At the end of the reaction chain, the nucleus is a thermally
equilibrated system, characterized by its excitation energy.
This system can ``evaporate'' nucleons, fragments,
or $\gamma$ rays, or can even fission, to dissipate the residual excitation.

Neutron emission is favoured over charged particle emission, due to the
Coulomb barrier, expecially  for medium-heavy
nuclei. Moreover, the excitation energy is higher
in heavier nuclei due to the larger
cascading chances and to the larger number of primary collisions in the Glauber
cascade at high energies. The level density parameter $a\approx A/8$~MeV
$^{-1}$ is higher too, thus the average neutron energy is smaller.
Therefore, the neutron multiplicity is higher for heavy
nuclei than for light ones.

The \FLUKA{} evaporation module is based on the standard Weisskopf-Ewing
formalism~\cite{Weisskopf}.
Latest improvements~\cite{FLUKA} are represented by: i)~adopted
state density expression \mbox{$\rho \propto \exp{(2\sqrt{aU})}/U^\frac{5}{4}$}
(where $U$ is the emitting nucleus excitation energy),
ii)~no Maxwellian approximation for energy sampling,
iii)~competition with $\gamma$ emission,
iv)~sub-barrier emission.
Neutron and proton production are marginally affected, while residual
nuclei production and alpha emission are nicely improved.

For light residual nuclei, where the excitation energy may
overwhelm the total binding energy, a statistical fragmentation
(Fermi Break-up) model is more appropriate
(see~\cite{FLUKA,Trieste,ourZPC1} for the \FLUKA{} implementation).

The evaporation/fission/break-up processes represent the last stage of a
nuclear interaction and are responsible for the exact nature of the
residuals left after the interactions. However, for a coherent
self-consistent model, the mass spectrum of residuals is highly constrained
by the excitation energy distribution found in the slow stages, which in
turn is directly related to the amount of primary collisions and following
cascading which has taken place in the fast stages.
\section{Nucleus--nucleus collisions \label{SCAAcoll}}
The \FLUKA{} implementation of suitable models for heavy ion nuclear
interactions has reached an operational stage. At medium/high energy
(above a few GeV/n) the \DPMJET{} model is used
as described in Subsection~\ref{SSC:Dpmjet}.
The major task of incorporating heavy ion
interactions from a few GeV/n down to the threshold for inelastic
collisions is also progressing and promising results have been
obtained using a modified version of the \RQMD-2.4 code (see
Subsection~\ref{SSC:RQMD}).

\subsection{The \FLUKA{} - \DPMJET{} interface \label{SSC:Dpmjet}}
\DPMJET-II.53 \cite{Ranft}, a Monte Carlo model for sampling
 h--h, h--A and nucleus-nucleus (A--A) collisions at
 accelerator and cosmic ray energies (E$_{lab}$ from 5-10~GeV/n up to
 10$^{9}$-10$^{11}$ GeV/n) was adapted and interfaced to \FLUKA{}.
\FLUKA{} implements \DPMJET-II.53 as an event generator to simulate
 A--A interactions exclusively. \DPMJET{} (as well as the \FLUKA{}
 high energy h--A generator) is based on the Dual Parton Model in
 connection with the Glauber formalism. The implementation of \DPMJET{} is
 also considered a possible, future option to extend the \FLUKA{} energy limits
 for hadronic simulations in general.

 Internally, \DPMJET{} uses Glauber impact parameter distributions per
 projectile--target combination.  These are either computed during
 initialization of the program or can be processed and output in a
 dedicated run of \DPMJET{} in advance. The computations are CPU intensive for
 heavier colliding nuclei and it would not be practical to
 produce the required distributions repeatedly while processing full showers in
 \FLUKA{}. Therefore, a procedure was devised to efficiently provide
 pre-computed impact parameter distributions for a complete matrix of
 projectile--target combinations up to a mass number A=246 over the
 whole available energy range~\cite{Empl}.

 \FLUKA{} requires A--A reaction cross sections internally in order to select
 A--A interactions appropriately. Hence, a complete matrix of A--A
 reaction cross sections was prepared along with the Glauber impact
 parameter distributions. Owing to the well established validity of the
 Glauber formalism, these cross sections can be safely applied down to a
 projectile kinetic energy $\approx$1~GeV/n.

 \DPMJET{} is called once per A--A interaction. A list of final state particles
 is returned by \DPMJET{} for transport to \FLUKA{}, as well as up to two excited
 residual nuclei with their relevant properties. De-excitation
 and evaporation of the excited residual nuclei is performed by calling the
 \FLUKA{} evaporation module.

 Work to interface \DPMJET-III \cite{Roesler} is in progress.

\subsection{The \FLUKA{} - \RQMD{} interface \label{SSC:RQMD}}
\begin{figure}[hbt]
\begin{center}
\epsfig{file=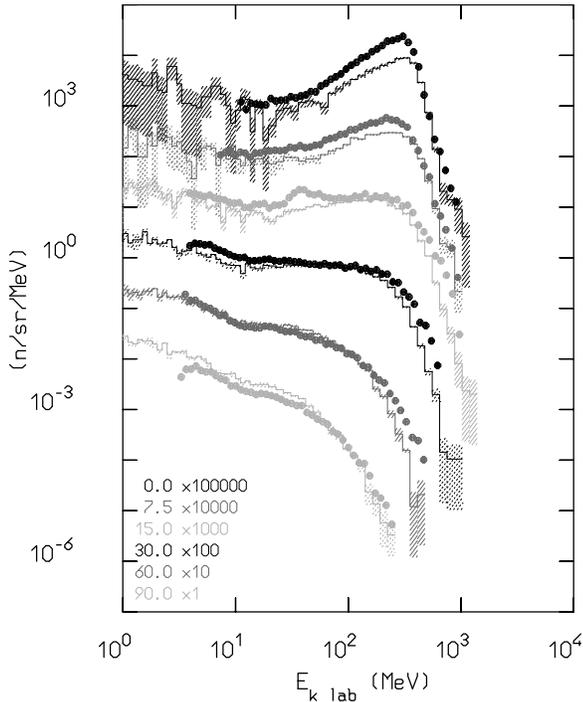,width=0.45\textwidth%
,bbllx=18pt,bblly=74pt,bburx=560pt,bbury=732pt,clip=}
\caption{
Double differential neutron yield by 400~MeV/n~Ar~ions
on thick Al targets. Data are shown for six different
laboratory emission angles, with the most forward on top:
histograms are \FLUKA{} results, dots experimental data~\cite{Kuro}.
\label{fig:400ar}
}
\end{center}
\end{figure}
\begin{figure}[t]
\begin{center}
\epsfig{file=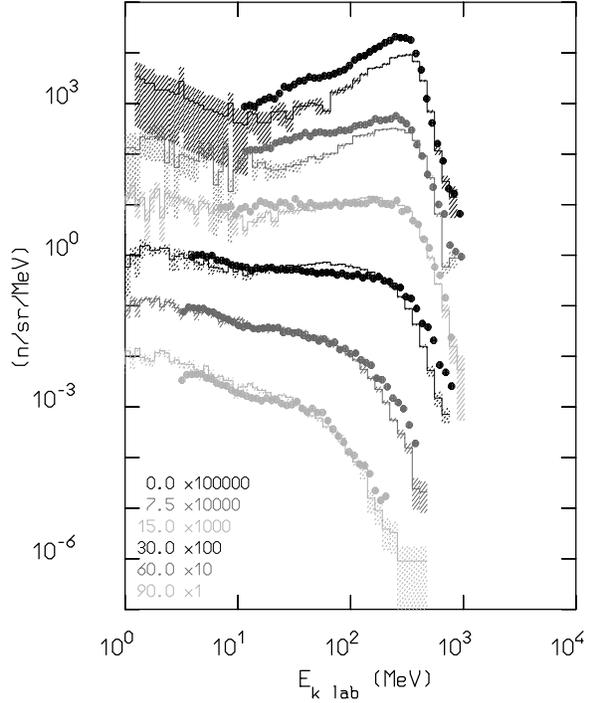,width=0.45\textwidth%
,bbllx=18pt,bblly=74pt,bburx=560pt,bbury=732pt,clip=}
\caption{
The same as fig.~\ref{fig:400ar} for Fe ions.
\label{fig:400fe}
}
\end{center}
\end{figure}
\begin{figure}[h]
\begin{center}
\epsfig{file=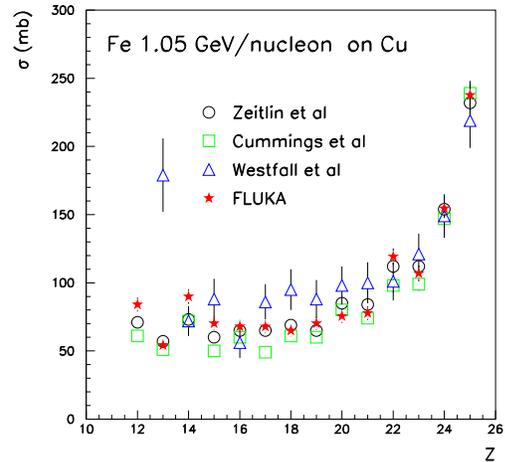,width=0.45\textwidth%
,bbllx=-0pt,bblly=40pt,bburx=560pt,bbury=560pt}
\caption{
Fragment charge cross sections for 1.05~GeV/n Fe ions on Cu.
Stars \FLUKA, circles~\cite{Zeit}, squares (1.5~GeV/n)
~\cite{Cumm}, triangles (1.88~GeV/n)~\cite{West}.
\label{fig:x1gevfecu}
}
\end{center}
\end{figure}
\begin{figure}[t]
\begin{center}
\epsfig{file=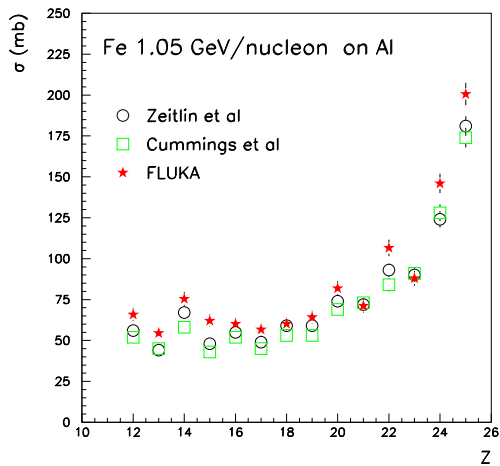,width=0.45\textwidth%
,bbllx=-0pt,bblly=40pt,bburx=560pt,bbury=560pt}
\caption{
The same as fig.~\ref{fig:x1gevfecu} for 1.05~GeV/n Fe ions on Al.
\label{fig:x1gevfeal}
}
\end{center}
\end{figure}
Quantum Molecular Dynamics (QMD) approaches are a viable solution for
A--A reactions. They represent an improvement over classical
INC codes, thanks to their dynamic modelling of the
nuclear field among nucleons during the reaction.
The treatment of
individual two-body scattering/interactions is usually based on similar
approaches for INC and QMD codes.
Unfortunately, initialization of the projectile and target nuclear states
is often difficult and their relativistic extension somewhat problematic.

The \RQMD-2.4~\cite{Greiner,Sorge} is a  relativistic QMD model
which has been applied successfully to relativistic A--A particle production
over a wide energy range, from $\approx$ 0.1~GeV/n up to several hundreds
of GeV/n. The high energy A--A  part in \FLUKA{} is
already successfully covered by \DPMJET{}. For energies below a few GeV/n
several models are under development, either new (see Section \ref{SCperspe}),
or based on the extension of the present intermediate energy h--A model of
\FLUKA{}.
However, a \RQMD-2.4 interface has been developed meanwhile to enable
\FLUKA{} to treat ion interactions from $\approx$100~MeV/n up to 5~GeV/n
where \DPMJET{} starts to be applicable.

Several important issues had to be addressed. \RQMD{} does not identify
nuclei in the final state. Hence, no low energy de-excitation (evaporation,
fragmentation\ldots) is performed for neither the excited projectile
nor the target residues. This is highly problematic, particularly
for the projectile residue, since its de-excitation usually gives rise to
the highest energy particle production in the laboratory frame.
Serious energy non-conservation issues are also affecting the code,
particularly when run in full QMD mode (\RQMD{} can be run both in full QMD
mode and in the so called ``fast cascade'' mode where it behaves as an
INC code).
Therefore a meaningful calculation of residual excitation energies was
impossible with original code.

The adopted solution was to modify the code, reworking
the nuclear final state out of the available information on spectators,
correlating the excitation energy to the actual depth of holes
left by hit nucleons.
Finally, the remaining energy-momentum conservation issues were resolved
taking into account experimental binding energies, as in all
other \FLUKA{} models.
After these improvements a meaningful excitation energy could be
computed and the \FLUKA{} evaporation model (see Subsection \ref{SSCevap})
was used to produce the particles of low energy (in their respective rest frames)
emitted by the excited projectile and target residues.

Examples of the performances of \FLUKA{} when run with the modified
\RQMD-2.4 code are presented in figs.~\ref{fig:400ar},\ref{fig:400fe},
\ref{fig:x1gevfecu}, and~\ref{fig:x1gevfeal} and compared with experimental
data, for double differential neutron production at 400~MeV/n and fragment
production at 1.05~GeV/n.

\section{Perspectives \label{SCperspe}}
\begin{figure}[htb!]
\begin{center}
\epsfig{file=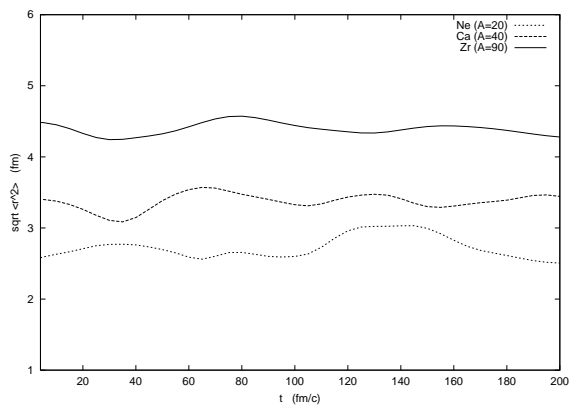,width=0.45\textwidth}
\caption{
Root mean square radii (fm)
of selected ${^{20}\mathrm{Ne}}$, ${^{40}\mathrm{Ca}}$,
${^{90}\mathrm{Zr}}$
initial nuclear configurations versus time (fm/c).
\label{fig:raggi}
}
\end{center}
\end{figure}

\begin{figure}[htb!]
\begin{center}
\epsfig{file=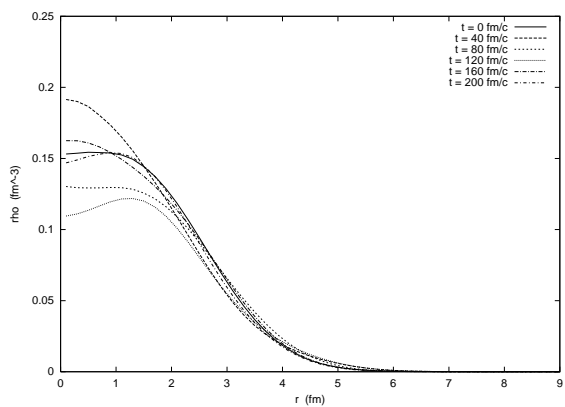,width=0.45\textwidth}
\caption{
Radial spatial density profile of the same
${^{20}\mathrm{Ne}}$ configuration shown in fig.~\ref{fig:raggi},
at the beginning of the time
evolution (solid line) and at subsequent times ($\Delta t$ = 40 fm/c).
\label{fig:rhoNe148}
}
\end{center}
\end{figure}

A new QMD code is also being developed from scratch to describe A--A
collisions taking into account both the effect of stochasting
scattering between all the nucleons involved and the effect
of the nuclear potential which acts on each nucleon. Protons and
neutrons are described as gaussian wave packets of fixed widths;
the total nuclear wave function is approximated by the product of single
nucleon wave functions.
Regarding the Hamiltonian, each group of QMD model developers
makes its particular choices
(see, for instance, refs.~\cite{Greiner,Aiche,Souza,Niita,Wang,Papa});
our starting point is a
non-relativistic phenomenological potential,
based on Skyrme interaction, supplemented by surface and symmetry terms;
we add also the electromagnetic repulsion between protons,
crucial to determine low-energy nuclear trajectories.
In principle one can build an Hamiltonian as sophisticated
as desired to improve the nuclear physical description;
in practice however one has to meet CPU time requirements.

The parameters of the model are fixed in order to reproduce as accurately as possible
the observed nuclear ground state properties. We emphasize that this
result can be achieved only approximatively, because
one has to describe with only a few parameters a great variety of nuclei,
ranging from the lightest to the heaviest ones.
For this purpose a gaussian width increasing
with increasing mass number turns out to be useful,
as suggested for istance in \cite{Wang}.
On the other hand, taking a gaussian width which differs from nucleon
to nucleon and evolves in time, as done for istance in \cite{Maruy},
is practicable only for light nuclei; for the intermediate mass and
heavy ones the required CPU time becomes higher and higher.

We underline that QMD codes are based on Monte Carlo simulations:
the cross-sections for A--A interactions are obtained as mean
values from hundreds and hundreds of events, each of which
should involve different starting nuclear configurations. In practice,
it turns out that one can simulate a wide variety of different events
with only a few initial configurations, randomly rotated.

Reasonable initial configurations can be chosen and stored. 
The time evolution of root mean square radii for selected
${^{20}\mathrm{Ne}}$, ${^{40}\mathrm{Ca}}$,
${^{90}\mathrm{Zr}}$  configurations is shown in fig.~\protect\ref{fig:raggi};
the rms radii slightly oscillate
with time because the nucleons are not frozen inside the
nucleus: each of them moves in the potential well originated
from all the others because of its Fermi momentum, different from zero.
Note that QMD initial configurations are not classical Hamiltonian minima,
which would break the Pauli principle~\cite{Niita}.
One has to make sure that
no nucleons escape at least for a time of the order of that required
to A--A collisions to take place ($\sim$ hundreds of fm/c);
only configurations which do not originate this spurious
emission are selected and stored for subsequent simulations.
.

The stored configurations should also provide
reasonable values for density and momentum distributions.

The radial spatial density profile is
plotted in fig.~\protect\ref{fig:rhoNe148} for the same
${^{20}\mathrm{Ne}}$ nucleus whose root mean square
radius evolution is shown in fig.~\protect\ref{fig:raggi};
each curve refers to a different time during evolution.
One can see that, as well as the radius, also the density oscillates
around its typical mean value.
Similar plots can be obtained for the time evolution of the momentum
distribution.

It is planned to
couple the dynamical nuclear evolution predicted by our QMD,
which gives the description of the first stage of the reaction,
with the \FLUKA{} preequilibrium module, to describe the deexcitation of
the fragments formed and to study deeply the fragmentation process,
implementing suitable models.

Moreover, a promising task is represented by the coupling of \FLUKA{}
with a Monte Carlo code~\cite{Riva} developed at Milan University and
based on Boltzmann Master Equation theory,
as a tool to treat ion--ion interactions below 100~MeV/n.

\begin{acknowledgments}
This work was partially supported under NASA Grant NAG8-1658,
ASI contract 1/R/320/02, and EC contract FIGH-CT1999-00005.
\end{acknowledgments}


\end{document}